\documentclass[runningheads]{llncs}
\usepackage{graphicx}
\usepackage{amsmath}

\usepackage{stfloats}

\usepackage{amsfonts}
\usepackage{graphicx}
\usepackage{booktabs}
\usepackage{float}

\usepackage{multirow}

\begin{document}
	%

	
		\title{Convolutional 3D to 2D Patch Conversion for Pixel-wise Glioma Segmentation in MRI Scans}
	
	%
	\titlerunning{Convolutional 3D to 2D Conversion for Pixel-wise Glioma Seg. in MRI Scans}
	
	\author{Mohammad~Hamghalam\inst{1,2}\orcidID{0000-0003-2543-0712} \and
		Baiying~Lei\inst{1} \and
		Tianfu~Wang\inst{1}}
	\authorrunning{M. Hamghalam et al.}
	%
	\institute{National-Regional Key Technology Engineering Laboratory for Medical Ultrasound, Guangdong Key Laboratory for Biomedical Measurements and Ultrasound Imaging, School of Biomedical Engineering, Health Science Center, Shenzhen University, Shenzhen, China, 518060. \and
		Faculty of Electrical, Biomedical and Mechatronics Engineering, Qazvin Branch, Islamic Azad
		University, Qazvin, Iran.\\
		\email{m.hamghalam@gmail.com}\\
		\email{\{leiby,tfwang\}@szu.edu.cn}}
	
	%
	%
	%
	%
	\maketitle              
	\begin{abstract}
		Structural magnetic resonance imaging (MRI) has been widely utilized for analysis and diagnosis of brain diseases. Automatic segmentation of brain tumors is a challenging task for computer-aided diagnosis due to low-tissue contrast in the tumor subregions. To overcome this, we devise a novel pixel-wise segmentation framework through a convolutional  3D to 2D MR patch conversion model to predict class labels of the central pixel in the input sliding patches. Precisely, we first extract 3D patches from each modality to calibrate slices through the squeeze and excitation (SE) block. Then, the output of the SE block is fed directly into subsequent bottleneck layers to reduce the number of channels. Finally, the calibrated 2D slices are concatenated to obtain multimodal features through a 2D convolutional neural network (CNN) for prediction of the central pixel. In our architecture, both local inter-slice and global intra-slice features are jointly exploited to predict class label of the central voxel in a given patch through the 2D CNN classifier. We implicitly apply all modalities through trainable parameters to assign weights to the contributions of each sequence for segmentation. Experimental results on the segmentation of brain tumors in multimodal MRI scans (BraTS'19) demonstrate that our proposed method can efficiently segment the tumor regions.
		
	\end{abstract}
	\section{Introduction}    
	\label{sec:intro}
	
\begin{figure}[tbp]
		\centering    
		\includegraphics[width=8.8cm]{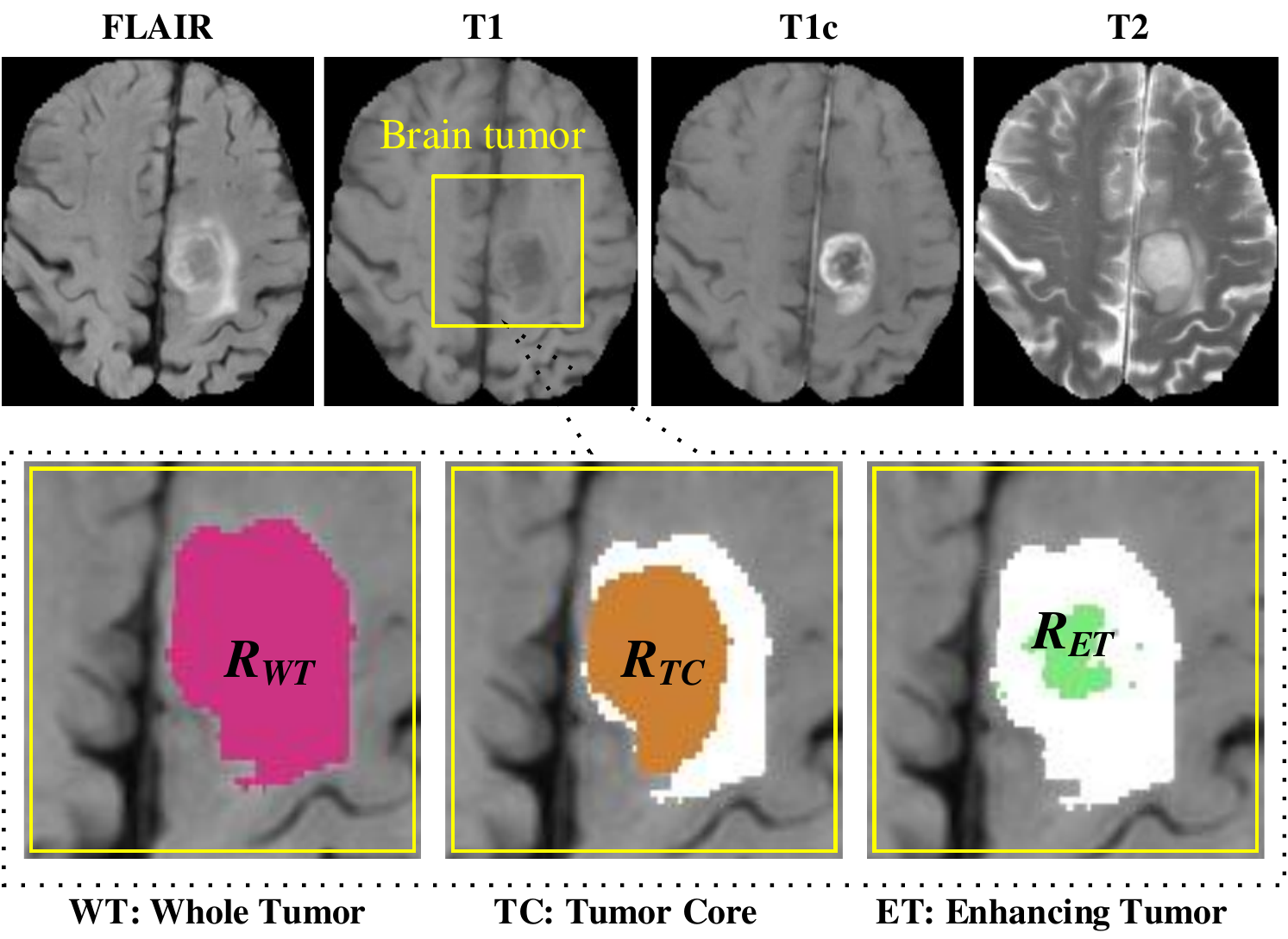}  
		\vspace{-0.3cm}
		\caption{Structural MRI provides a non-invasive method to determine abnormal changes in the brain for clinical purpose. Four MRI modalities (FLAIR, T1, T1c, and T2) along with brain lesion: WT (all internal parts), TC (all except edema), and ET (enhancing tumor).} 
		
		\label{fig:modality}
	\end{figure}

	Among brain tumors, glioma is the most aggressive and prevalent tumor that begins from the tissue of the brain and hopefully cannot spread to other parts of the body. Glioma can be classified into low-grade glioma (LGG) and high-grade glioma (HGG). LGGs are primary brain tumors and usually affect young people compared to HGGs.  Multimodal MR sequences comprised of FLAIR, T1, T1c, and T2 are usually used to segment internal parts of the tumor, i.e., whole tumor (WT), tumor core (TC), and enhancing tumor (ET) as depicted in Fig. \ref{fig:modality}. Since the shape and location of tumors are unpredictable, it is difficult to identify exactly type of brain tumor by studying the brain scans. On the other hand, the low tissue contrast in the lesion regions makes the tumor segmentation a challenging task. Moreover, manual annotation of these tumors is a time-consuming and often biased task. Thus, automatic segmentation approaches are a crucial task in diagnosis, analysis, and treating plane.
	
	\begin{figure*}[ht]

		\includegraphics[width=12cm]{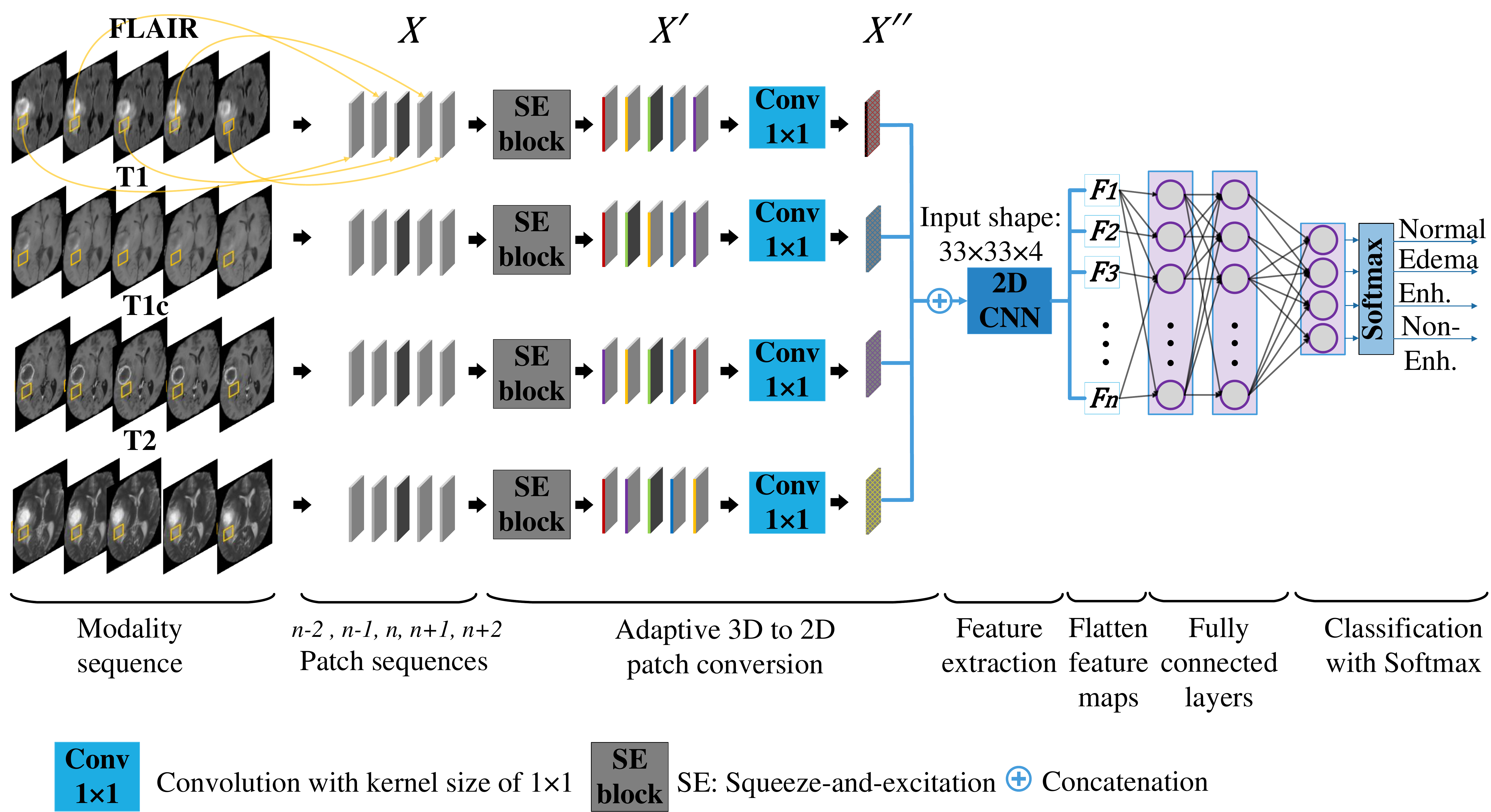}
		
		\caption{Illustration of the proposed 3D to 2D patch conversion model for the pixel-wise segmentation of glioma. The 3D to 2D model includes the SE block and $1\times1$ convolution as the bottleneck layer. The input data are 3D multimodal image patches, and the outputs are four types of clinical subregions, i.e., edema, non-enhancing, enhancing, and healthy tissue.}
		\label{fig:framework}
		
	\end{figure*}

	 Many segmentation methods have been proposed to segment tissue of interest based on traditional \cite{HamghalamAutomatic2009,Hamghalam2009,Soleymanifard2019,Soleimany2017} and modern machine learning methods \cite{Hamghalam2019Brain} in medical application. The brain tumor segmentation methods \cite{Hatami2019,Najrabi2018} can be roughly categorized into the pixel-wise \cite{pereira2016brain,Havaei2017} and region-wise \cite{isensee2017brain,Guotai2017,pereira2018adaptive,shen2017boundary,le2018deep} techniques. The former predicts only the central pixel of each input patch while the latter predicts labels of the most pixels inside the input patches.  
	The region-wise methods are usually based on 3D \cite{Guotai2017,isensee2017brain,pereira2018adaptive}  and 2D \cite{shen2017boundary,le2018deep} fully convolutional networks (FCNs). 
	Wang \textit{et al.} \cite{Guotai2017} applied the cascaded framework with three stages to segment WT, TC, and ET on each stage, respectively.
	Isensee \textit{et al.} \cite{isensee2017brain} employed a U-Net-like architecture \cite{ronneberger2015u} that was trained on the BraTS training dataset \cite{Menze2015Brain,bakas2017advancing} along with a private clinical dataset with some augmentations.
	In another work, Pereira \textit{et al.} \cite{pereira2018adaptive} introduced two new blocks to extract discriminative feature maps: recombination-recalibration (RR) and segmentation squeeze-and-excitation (SegSE) blocks.
	
	In 2D structures, Shen \textit{et al.} \cite{shen2017boundary} utilized a multi-task FCN framework to segment tumor regions. Additionally, Le \textit{et al.} \cite{le2018deep} introduced deep recurrent level set (DRLS) based on VGG-16 with three layers: convolutional, deconvolutional, and LevelSet layer. 
	In the pixel-wise networks \cite{pereira2016brain,Havaei2017}, the authors established the 2D CNN-based model to predict a single class label for the central pixel in the 2D multimodal patches. However, the intra-slice features are not used in their segmentation frameworks.

	Although the 3D FCN models can capture 3D information from MRI scans, 3D architectures are too computationally expensive because of the complicated network structure, including the 3D kernels, 3D input patches, and input dimensions. Notably, the size of the image patches is the most notable memory factor in convolutional nets, especially in the multimodal BraTS scans with four sequences. In the case of multimodal 3D scans, we have 5-dimensional tensors, including batch size, width, length, depth, and the number of modality concatenation. These tensors require much more memory for training and testing compared to 2D FCN.

	The focus of the current study is to develop a 3D to 2D conversion network for the pixel-wise segmentation. The conversion block employs squeeze-and-excitation (SE) block to adaptively calibrate slices in the input patch sequence by explicitly modeling the interdependencies between these slices. The bottleneck layer is applied to encode the 3D patches to 2D ones to decrease the number of input channel to the following feature extraction block. We use multimodal 2D output patches for segmentation through the 2D-CNN network. Particularly, we utilize the 3D feature between consecutive slices while using convolutional layers with 2D kernels in our framework.
	The rest of our paper is organized as follows. In Section \ref{sec:Method}, we describe 3D to 2D conversion method. Section \ref{sec:Experimental_Results} explains the databases used for evaluation and experimental results. Some conclusions are drawn in Section \ref{sec:Conclusion}. 
	\section{Method}
	\label{sec:Method}
	Our goal is to segment an input MR volume, $\mathbf{I}\in {{\mathbb{R}}^{H\times W\times D}}$, according to manual labels $\mathbf{S}\in {{\{1,2,...,c\}}^{H\times W\times D}}$, where $c$ is the number of output classes. Also $H$, $W$, and $D$ are the spatial height, width, and depth, respectively. Let ${{\mathbf{x}}}\in {{\mathbb{R}}^{\omega \times \omega \times L}}$ denotes the cropped 3D input patch on the central voxel, ${{x}^{\frac{\omega}{2},\frac{\omega}{2},\frac{L}{2}}}$. We need to predict the label of central voxels in each extracted 3D patch via 2D-CNN network.
	Fig. \ref{fig:framework} demonstrates an overview of the proposed method. We first introduce the adaptive 3D to 2D conversion module, and then 2D-CNN architecture will be discussed.

	\subsection{Convolutional 3D to 2D Patch Conversion}
	\label{sssec:3D-to-2D}
	
	We extend the SE block \cite{hu2018squeeze} to deal with the calibration of input 3D patches. Our model squeezes the global spatial information in each slice by computing average in each slice as:
	\begin{equation}\label{eq1}
	\begin{aligned}
	{{{z}}_{l}}={{F}_{sq}}({{\mathbf{x}}_{l}})=\frac{1}{\omega\times \omega}\sum\limits_{i=1}^{\omega}{\sum\limits_{j=1}^{\omega}{{\mathbf{x}}_{l}(i,j).}} 
	\end{aligned}
	\end{equation}
	where ${{{z}}_{l}}$ is the global embedded information in the slice of $l$.
	The second operation called ‘excitation’ is applied to capture slice-wise dependencies with a sigmoid ($\sigma $) and ReLU ($\delta $) activation, respectively. Thus we have:
	\begin{equation}\label{eq2}
	\begin{aligned}
	\mathbf{u}={{F}_{ex}}(\mathbf{z},W)=\sigma ({{W}_{2}}\delta ({{W}_{1}}\mathbf{z}))
	\end{aligned}
	\end{equation}
	where ${{W}_{1}}\in {{\mathbb{R}}^{r\times {{w}^{2}}}}$ and ${{W}_{2}}\in {{\mathbb{R}}^{{{w}^{2}}\times r}}$ are the weight matrices of two fully-connected with reduction ratio $r$.  
	At last, the scalar ${u}_{l}$  and input slice $\mathbf{{x}}_{l}$  are multiplied to obtain the calibrated 3D patch, $\mathbf{{x}'}\in {{\mathbb{R}}^{\omega \times \omega \times L}}$.
	
	Our bottleneck layer is a block that contains one convolutional layer with the kernel size of $1 \times 1$ to represent calibrated 3D slices as 2D with nonlinear dimensionality reduction, $\mathbf{x}''$. Each 2D patch thus forms a 3D-like representation of a part ($n$ consecutive slices) of the MR volume. This model allows incorporating some 3D information while bypassing the high computational and memory requirements of the 3D CNN.
	
	\subsection{Classifier Block for Pixel-wise Prediction}
	\label{sssec:Classifier block}
	
	The output slices from four 3D to 2D blocks are concatenated and fed into classifier block to predict the label of voxel where is located at the center of its cropped patch. The proposed network allows jointly capturing contextual features from FLAIR, T1, T1c, and T2 modality. For feature extraction, we rely on CNN block to learn from ground truth scores.  Our feature extractor consists of two levels of $3\times3$ convolutions along with max-pooling layers. The number of kernels in each level is 32, 32, 32, 64, 64, and 64, respectively. The fully-connected layers are composed of 64 and 32 hidden neurons, respectively, followed by the final Softmax layer. 
	Finally, we optimize cross-entropy loss between the predicted score, ${{F}_{seg}}({{\mathbf{x}}^{FLAIR}},{{\mathbf{x}}^{T1}},{{\mathbf{x}}^{T1c}},{{\mathbf{x}}^{T2}};\mathbf{W})$, and the ground truth label, ${{s}^{\frac{\omega}{2},\frac{\omega}{2},\frac{L}{2}}}$, with ADADELTA optimizer \cite{Matthew2012} as:   
	\begin{equation}\label{eq3}
	\underset{\mathbf{W}}{\mathop{\arg \min }}\,-\sum\limits_{i}^{c}{{{s}^{\frac{\omega }{2},\frac{\omega }{2},\frac{L}{2}}}.\log ({{F}_{seg}}({{\mathbf{x}}^{FLAIR}},{{\mathbf{x}}^{T1}},{{\mathbf{x}}^{T1c}},{{\mathbf{x}}^{T2}};\mathbf{W}))}
	\end{equation}
	where $c$ is the class number and $\mathbf{W}$ is the trainable parameter of the model. 
	
	\section{Experimental Results}
	\label{sec:Experimental_Results}
	
	\subsection{Implementation Details}
	\label{sec:Implementation}
	We implement the proposed method using the KERAS and TensorFlow with 12GB NVIDIA TITAN X GPU. We have experimentally found that volumes of seven have the best compromise between accuracy and complexity. Thus, the input MR volumes are partitioned into $33\times33 \times 7$ patches at the center of each label, then the concatenated patches from four modalities are considered as training data. For efficient training and class imbalance in brain tumor,  we perform augmentation in the number of patches for the small sample size classes.  The model is trained using the ADADELTA \cite{Matthew2012} optimizer 
	(learning rate = 1.0, $\rho=0.95$, epsilon=1e-6) and cross-entropy as the loss function. Dropout is employed to avoid over-fitting during the training process ($p_{drop}=0.5$).
	\subsection{Datasets}
	\label{sec:Dataset}
	The performance of the proposed pixel-wise method is evaluated on BraTS \cite{Bakas2018_Identifying,Bakas2017_Radiomic_2,Bakas2017_Radiomic_1,bakas2017advancing,Menze2015Brain} dataset to compare with other segmentation methods based on the pixel. BraTS'13 contains small subjects, i.e., 30 cases for training and 10 cases for the Challenge. We additionally evaluate the proposed technique on BraTS'19, which has two publicly available datasets of multi-institutional pre-operative MRI sequences: Training (335 cases) and Validation (125 cases). Each patient is contributing $155\times240\times240$ with four sequences: T1, T2, T1c, and FLAIR. In BraTS'19, it identifies three tumor regions: non-enhancing tumor, enhancing tumor, and edema. Evaluation is performed for the WT, TC, and ET. The evaluation is assessed by the SMIR\footnote{ https://www.smir.ch/BRATS/Start2013} and CBICA IPP\footnote{https://ipp.cbica.upenn.edu} online platforms.
	Metrics computed by the online evaluation platforms in BraTS'19 are Dice Similarity Coefficient (DSC) and the 95th percentile of the Hausdorff Distance (HD95), whereas, in BraTS'13, the online platform calculates DSC, Sensitivity, and Positive Predictive Value (PPV). DSC is considered to measure the union of automatic and manual segmentation. It is calculated as $DSC = \frac{2TP}{FP+2TP+FN}$ where TP, FP, and FN are the numbers of true positive, false positive, and false negative detections, respectively.

	\subsection{Segmentation Results on BRATS'13}
	
	\subsubsection{Ablation Study}
	\label{sec:Ablation Study}
	
	To investigate the effect of the proposed adaptive 3D to 2D block, we perform experiments with and without considering the 3D to 2D block. For the latter, we directly apply multimodal 3D volume into the 3D plain CNN model. We train both models with the 320K patch for an equal number of the patch in each group and validate on ten unseen subjects. Also, Dropout is employed to avoid over-fitting during the training process ($ p_{drop}=0.5 $). As presented in Table \ref{tab:ablation}, the results with 3D to 2D block increase the accuracy of segmentation in terms of standard evaluation metrics compared to the 3D baseline.  
	
	\begin{table*}[!t]
		\centering
		\caption{Impact of the 3D to 2D conversion block in segmentation: we perform experiments using the same setting to evaluate performance with and without proposed block.} 
		\label{tab:ablation}
		
		\begin{tabular}{@{\extracolsep{\fill}}ccccccccc}
			\toprule[1pt] \midrule[0.3pt]
			\rule[-1ex]{0pt}{2.5ex}
			\multirow{3}{*}{\textbf{Model}} && \multicolumn{3}{c}{\textbf{DSC}} && \multicolumn{3}{c}{\textbf{HD95 (mm)}  }  \\
			\cmidrule[\heavyrulewidth]{3-5} 
			\cmidrule[\heavyrulewidth]{7-9}
			&&\textbf{ET} &\textbf{WT} & \textbf{TC} && \textbf{ET} &\textbf{WT} & \textbf{TC}\\
			
			\midrule 
			With 3D to 2D block   		&& 83.15    &91.75&    92.35	&&    1.4	&    3.6	&1.4\\
			Without 3D to 2D block   	&& 80.21    &89.73&    88.44	&&    3.1	&   4.2	    &1.5\\

			\midrule[0.3pt]\bottomrule[1pt]
		\end{tabular}
	\end{table*}

	\subsubsection{Comparison with State-of-the-arts}	
	
	We also compare the performance of the proposed method with the well-known pixel-wise approach \cite{pereira2016brain,Havaei2017} and 2D region-wise ones \cite{shen2017boundary} on BraTS'13 Challenge. Table \ref{tab:cmp_others} shows DSC $(\%)$, Sensitivity, and  PPV for EN, WT, and TC, respectively. Moreover, it can be seen that the proposed method outperforms others in DSC for WT.

	\begin{table*}[htbp]
		\centering
		\caption{Comparison of proposed 3D to 2D method with others on BraTS'13 Challenge dataset.} 
		\label{tab:cmp_others}
		
		\begin{tabular}{@{\extracolsep{\fill}}ccccccccccccc}
			\toprule[1pt] 
			\rule[-1ex]{0pt}{2.5ex}
			\multirow{3}{*}{\textbf{Method}} && \multicolumn{3}{c}{\textbf{DSC}} && \multicolumn{3}{c}{\textbf{Sensitivity}}  && \multicolumn{3}{c}{\textbf{PPV}}  \\
			\cmidrule[\heavyrulewidth]{3-5} \cmidrule[\heavyrulewidth]{7-9} \cmidrule[\heavyrulewidth]{11-13}
			
			&&\textbf{EN} &\textbf{WT} & \textbf{TC} && \textbf{EN} &\textbf{WT} & \textbf{TC}&&\textbf{EN} &\textbf{WT} & \textbf{TC}\\
			
			\midrule 
			
			Shen \cite{shen2017boundary}        && 0.76  & 0.88 &    \textbf{0.83} &&    \textbf{0.81} & \textbf{0.90} & 0.81 && 0.73  &  0.87  & \textbf{0.87}  \\
			
			Pereira  \cite{pereira2016brain}    && \textbf{0.77}  &0.88 &    \textbf{0.83} &&    \textbf{0.81} & 0.89 & 0.83 && \textbf{0.74}  &  0.88  & \textbf{0.87}  \\
			
			Havaei  \cite{Havaei2017}           && 0.73  &0.88 &    0.79 &&    0.80 & 0.87 & 0.79 && 0.68  &  0.89 & 0.79   \\
			
			Proposed  method             && 0.74  & \textbf{0.89} &    0.80 &&    0.78 & 0.86 & \textbf{0.86} && 0.73 &   \textbf{0.92}  & 0.76   \\

			\midrule[0.1pt]
		\end{tabular}
		
	\end{table*}

\subsection{Segmentation Results on BRATS'19}	

One limitation of pixel-wise methods is the time complexity at inference time due to pixel by pixel prediction. Specifically, we have to process about 9M voxels per channel for each patient. Although we eliminate voxels with the value of zero in testing time, the pixel-wise prediction still needs longer time compared to region-wise ones. This issue limits our method for evaluation on BraTS'19 with 125 validation samples. To decrease the inference time, we use a plain 3D U-Net model to solely predict  WT as an initial segmentation, which further allows us to compute a bounding box concerning tumor region for our pixel-wise method. In this way, the segmentation of the internal part of the tumor area is performed inside the bounding box. The results in Table \ref{tab:segmentation_result} show that our method achieved competitive performance on automatic brain tumor segmentation. Results are reported in the online processing platform by BraTS'19 organizer.

	\begin{table}[!t]
		\centering
		\caption{DSCs and HD95 of the proposed method on BraTS'19 Validation set (training on 335 cases of BraTS'19 training set).} 
		\label{tab:segmentation_result}

			\begin{tabular}{@{\extracolsep{\fill}}c|cccc|cccc|cccc|ccc}
			\toprule[1pt] 
			\rule[-1ex]{0pt}{2.5ex}
			\multirow{3}{*}{\textbf{}} & \multicolumn{3}{c}{\textbf{Dice}} && \multicolumn{3}{c}{\textbf{Sensitivity}  }  && \multicolumn{3}{c}{\textbf{Specificity}  } && \multicolumn{3}{c}{\textbf{HD95 (mm)}  } \\
			\cmidrule[\heavyrulewidth]{2-4} 
			\cmidrule[\heavyrulewidth]{6-8}
			\cmidrule[\heavyrulewidth]{10-12}
			\cmidrule[\heavyrulewidth]{14-16}
			&\textbf{ET} &\textbf{WT} & \textbf{TC} && \textbf{ET} &\textbf{WT} & \textbf{TC}&& \textbf{ET} &\textbf{WT} & \textbf{TC}&& \textbf{ET} &\textbf{WT} & \textbf{TC}\\
			
			\midrule 
			Mean 			 	&72.48	  &89.65	&79.56	&& 73.25	& 90.60	&79.57	&&99.87		& 99.45	& 99.69		&& 5.4	& 7.8	& 8.7  \\
			\midrule 		 	                                        
			Std.      		 	&29.47	  &8.968	&21.62	&& 26.61	& 08.91	&24.77	&&0.23.5	& 0.58	& 0.36		&& 9.2	& 15.5	& 13.5 \\
			\midrule 		 	                                        
			Median         	 	&84.46	  &92.19	&89.17	&& 83.20	& 93.66	&91.14	&&99.94		& 99.64	& 99.82		&& 2.2	& 3.1	& 3.8  \\
			\midrule         	                                        
			25     quantile  	&70.99	  &88.31	&74.63	&& 67.73	& 87.83	&72.88	&&99.84		& 99.26	& 99.56		&& 1.4	& 2.0    & 2.0  \\
			\midrule         	                                        
			75     quantile  	&89.22	  &94.72	&93.39	&& 88.71	& 96.58	&96.04	&&99.98		& 99.81	& 99.93		&& 4.2	& 5.3	& 10.2 \\

			\midrule[0.1pt]
		\end{tabular}

	\end{table}

\begin{figure*}[tb]
		
		\includegraphics[width=12cm]{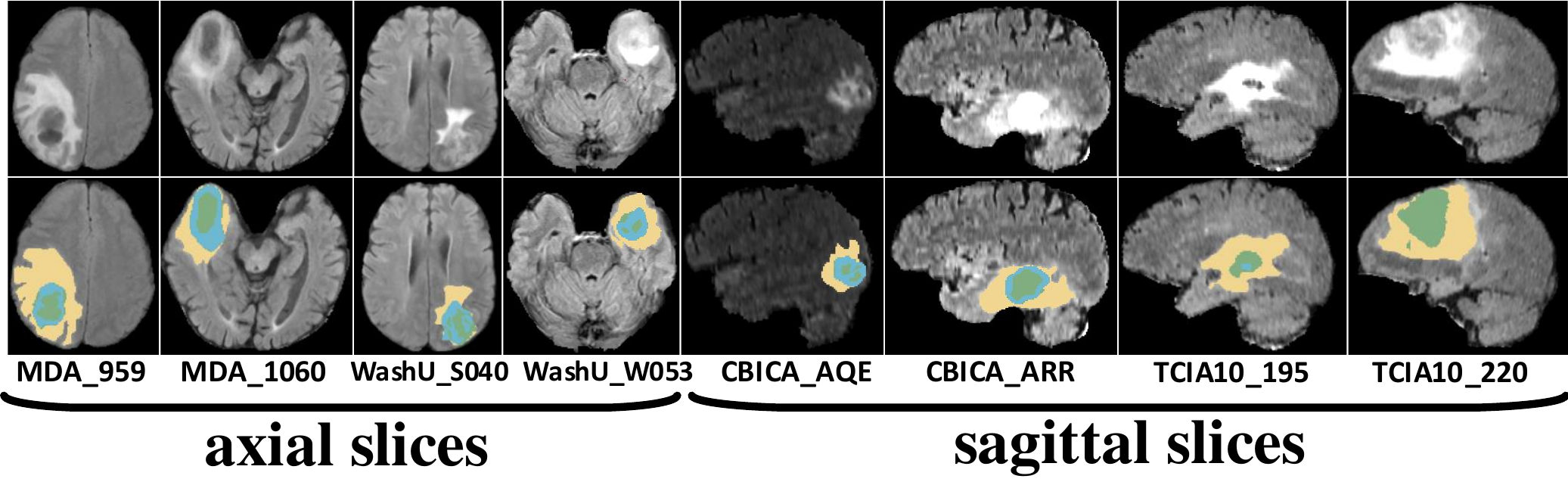}
		
		\caption{Segmentation results are overlaid on FLAIR axial and sagittal slices on BraTS'19 Validation Data. The yellow label is edema, blue color means enhancing tumor, and the green one presents the necrotic and non-enhancing tumor core. Each column displays one slice of different Subject IDs of BraTS'19.}
		\label{fig:seg_results}
		
	\end{figure*}

	Moreover, Fig. \ref{fig:seg_results} shows examples for glioma
	segmentation from validation slices of BraTS'19. For simplicity
	of visualization, only the FLAIR image is shown in the axial and sagittal view along with our segmentation results. The subject IDs in each column are related to the validation set.

	\section{Conclusion}
	\label{sec:Conclusion}
	
	This paper provides a framework that adaptively converts 3D patch into 2D to highlight discriminative pixels for the label prediction of central voxels. The converted 2D images are fed into the classifier block with 2D kernels for the predication. This conversion enables incorporating 3D features while bypassing the high computational and memory requirements of fully 3D CNN. We provided ablation study to examine the effect of our proposed conversion block on the segmentation performance. Results from the BraTs'13 and BraTS'19 dataset confirm that inter and intra-slice features effectively improve the performance while using 2D convolutional kernels. 
	Though pixel-wise methods have limitation in inference time, we can take advantage of pre-trained network for classification purpose through fine-tuning with MRI training set. Future works will concentrate on 3D to 2D patch conversion with an attention mechanism.
	
	\section{Acknowledgment}
\label{sec:Acknowledgment}
This work was supported partly by National Natural Science Foundation of China (Nos.61871274, 61801305, and 81571758), National Natural Science Foundation of Guangdong Province (No. 2017A030313377), Guangdong Pearl River Talents Plan (2016ZT06S220), Shenzhen Peacock Plan (Nos. KQTD2016053112 051497 and KQTD2015033016 104926), and Shenzhen Key Basic Research Project (Nos. JCYJ20170413152804728, JCYJ20180507184647636, JCYJ20170818142347 251, and JCYJ20170818094109846).

	%
	%
	%
	%
	\bibliographystyle{splncs04}
	\bibliography{Mybib_brats}

\end{document}